\documentclass{openjournal}
\usepackage{xcolor}
\usepackage{textgreek}
\usepackage[utf8]{inputenc}
\usepackage[english]{babel}
\usepackage{amsmath}
\let\oldAA\AA
\renewcommand{\AA}{\text{\normalfont\oldAA}}
\usepackage{subfigure}
\usepackage{booktabs}

\usepackage{hyperref}
\hypersetup{
    unicode, 
    colorlinks=true,
    linkcolor=linkcolor,
    citecolor=linkcolor,
    filecolor=linkcolor,
    urlcolor=linkcolor,
}
\usepackage{color,colortbl}
\definecolor{linkcolor}{rgb}{0.0,0.3,0.5}
\usepackage{tensind}
\tensordelimiter{?}
\DeclareGraphicsExtensions{.bmp,.png,.jpg,.pdf}
\usepackage{verbatim}
\usepackage[normalem]{ulem}
\usepackage{orcidlink}
\usepackage{soul}

\graphicspath{ {./figs/} }

\begin{document}
\title{Derivative-Aligned Anticipation of Forbush Decreases from Entropy and Fractal Markers}

\author{Juan D. Perez-Navarro$^{a,1}$\orcidlink{0009-0008-5689-314X} and D. Sierra-Porta$^{a,*}$\orcidlink{0000-0003-3461-1347}}
\email[Juan D. Perez-Navarro: ]{perezjuan@utb.edu.co}
\email[D. Sierra-Porta: ]{dporta@utb.edu.co}
\affiliation{Universidad Tecnológica de Bolívar. Escuela de Transformación Digital. Parque Industrial y Tecnológico Carlos Vélez Pombo Km 1 Vía Turbaco. Cartagena de Indias, 130010, Colombia}
\affiliation{$^{*}$ Corresponding author: dporta@utb.edu.co (D. Sierra-Porta)}

\begin{abstract}
We develop a feature-based framework to anticipate Forbush decreases (FDs) in one-minute neutron-monitor records by tracking sliding-window invariants from information theory, scaling, and geometry. For each station we compute marker time series—including Shannon, spectral, approximate and sample entropy; Lempel--Ziv complexity; correlation dimension; and Higuchi and Katz fractal dimensions—smooth them with an exponentially weighted moving average, and analyze their within-station standardized first differences. Timing is referenced to an operational alignment time $t_0$ defined as the minimum of the smoothed count first difference, and marker leads are reported in minutes ($\ell^*<0$ indicates anticipation). Station-level detectability is defined on a pre-$t_0$ window using a robust $z$-score detector with bilateral threshold and persistence, requiring neither cross-correlation nor hypothesis testing.

We apply the pipeline to two FD episodes with broad station coverage (2023-04-23 and 2024-05-10; 28 stations each). Across events, a compact CORE panel exhibits consistently high detection rates and predominantly negative lead distributions, with median leads of order several hours depending on the invariant and event. Lead dispersion across stations is substantial (interquartile ranges typically spanning a few hours), underscoring the value of station-wise criteria and distributional summaries rather than single-station inference. Representative marker trajectories confirm that early flagging corresponds to sustained pre-$t_0$ excursions in marker differences, not merely tabulated artifacts.

The approach is reproducible from open code, operates on native station units without cross-station homogenization, and is qualitatively stable to sensitivity sweeps of windowing, smoothing, and detector parameters. These results support derivative-aligned invariant panels as practical early-warning complements to amplitude-threshold methods in space-weather nowcasting.
\end{abstract}

% Write your keywords here
\begin{keywords}
    {Forbush decrease, space weather, neutron monitor, sliding-window invariants, entropy measures, fractal dimension}
\end{keywords}

%\maketitle

\section{Introduction}
\label{sec:intro}
Forbush decreases (FDs) are rapid depressions in galactic cosmic ray (GCR) intensity followed by a slower recovery, produced by transient heliospheric structures and interplanetary magnetic-field disturbances that modulate particle transport \citep{lockwood1971forbush}. They serve as operational diagnostics for space weather and as natural probes of heliospheric dynamics, because their onset, depth, and recovery encode information about large-scale magnetic turbulence, diffusion barriers, and evolving solar-wind structures. Beyond their scientific value, FDs co-vary with conditions that affect satellite operations, polar aviation, and ground-based infrastructure, motivating methods that can identify precursor signatures in neutron-monitor time series before the drop becomes manifest \citep{Balasisetal2023}. Evidence for long-range correlations and multifractality in both cosmic rays and heliospheric parameters further motivates the search for early-warning signals that go beyond amplitude changes in the raw count \citep{sierra2022fractal, sierra2022linking, sierra2024multifractal}.

A growing body of work shows that time-series invariants grounded in information theory, scaling, and geometry are useful descriptors of underlying dynamics. Highly comparative frameworks organize thousands of operations across disciplines, revealing empirical structure among methods and demonstrating that different measures are sensitive to complementary dynamical regimes \citep{fulcher2013highly, lubba2019catch22, amigo2022information}. In this study we treat such invariants as temporal markers: using a sliding window, each measure becomes its own time series that can be examined for \emph{temporal alignment} relative to the derivative-defined onset of the original GCR count signal, without relying on explicit cross-correlation analyses \citep{fulcher2014highly}.

The present work focuses on a set of topological–geometric and complexity measures that are both theoretically grounded and available in robust implementations. We consider Shannon, approximate, sample, permutation, and spectral entropies; Lempel–Ziv complexity; the Hurst exponent and its estimation via detrended fluctuation analysis (DFA); correlation dimension; and fractal dimensions due to Higuchi, Katz, and Petrosian. These measures quantify uncertainty, regularity, memory, roughness, and dimensionality at multiple scales. Comparative and tutorial studies highlight their interpretability and practical diversity, including guidance on parameterization and robustness in noisy, finite data \citep{unakafova2013efficiently, amigo2022information, raubitzek2021combining}. Classical nonlinear-dynamics estimators such as the information/correlation dimension provide geometric grounding for state-space structure \citep{Farmer1982}, while DFA-based scaling analysis links fluctuation growth to long-range dependence in a way that is compatible with weak nonstationarities commonly observed in environmental and space-physics series \citep{zhou2014multifault, Stolz2017AGS, keller2009standardized}.

Within the entropy family, permutation entropy offers a conceptually simple, ordinal-pattern view that is computationally efficient and has been used widely in biomedical and physical time series; extensions to non-uniform embeddings further improve sensitivity in systems with multiple characteristic scales \citep{unakafova2013efficiently, tao2018permutation}. Shannon and spectral entropies capture distributional and frequency-domain disorder, whereas approximate and sample entropy emphasize conditional irregularity in finite sequences, with well-known trade-offs in parameter choice and data-length requirements \citep{amigo2022information, unakafova2013efficiently}. Lempel–Ziv complexity provides a symbolization-based measure of algorithmic novelty growth and has proved effective as a complementary descriptor in large feature sets \citep{fulcher2013highly, lubba2019catch22}. Newer variants, such as SVD entropy, and methodological developments like complexity–entropy maps broaden the diagnostic palette for distinguishing dynamical regimes \citep{strydom2021svd, ribeiro2017characterizing, raubitzek2021combining}.

Fractal and scaling descriptors offer geometric and roughness-based perspectives. The Hurst exponent and the fractal dimension of the time-series graph are tightly related under self-similarity, yet can become complementary in intermittent settings typical of near-Earth space physics \citep{Balasisetal2023}. Practical estimation matters: comparative analyses report strong performance of Higuchi’s method and generalized Hurst estimators in self-affine settings, but also caution that sensitivity to transitions and windowed tracking requires careful methodological choices \citep{2016arXiv161106190K, 2013arXiv1310.3564H}. These considerations are directly relevant when one seeks temporal markers rather than single-snapshot characterizations.

Related work in space physics emphasizes the usefulness of complexity-science tools for diagnosing the near-Earth electromagnetic environment and for characterizing transitions across geomagnetic activity levels \citep{Balasisetal2023}. In cosmic-ray research specifically, multifractal analyses and cross-correlation studies between GCRs and heliospheric parameters have reported persistent correlations and scale-dependent structure consistent with diffusion and convection processes modulated by evolving magnetic turbulence \citep{sierra2024predicting, sierra2022fractal, sierra2025multifractal}. Complementary literature on feature-based time-series learning shows that entropy, fractal, and scaling features can be integrated into supervised and unsupervised workflows, improving detection and ranking of candidate precursors while providing a common ground for cross-station comparisons \citep{fulcher2013highly, fulcher2014highly, lubba2019catch22, raubitzek2021combining}. Methodological reviews focused on permutation, approximate, and sample entropy, as well as related complexity measures, provide practical insight into sensitivity to noise, embedding and window choices, and computational efficiency—considerations that are critical when working with neutron-monitor data at varying cadences \citep{unakafova2013efficiently, amigo2022information}. In parallel, comparative studies of entropy-like descriptors across fields (biomedicine, ecology, finance) illustrate their transferability and the value of multi-measure ensembles, from classical estimators to more recent quantities like bubble and PCA-based entropies \citep{Manisetal2021,Pandaetal2023, strydom2021svd,ribeiro2017characterizing}.

From the perspective of highly comparative analysis, large libraries of features such as \texttt{hctsa} and the reduced canonical subset \texttt{catch22} offer standardized, interpretable implementations spanning autocorrelation structure, fluctuation scaling, distributional deviations, and nonlinear descriptors \citep{fulcher2013highly, lubba2019catch22}. This ecosystem facilitates reproducible screening of candidate markers and systematic evaluation of redundancy, aiding the selection of a compact, physically meaningful set tailored to GCR data. Ordinal-pattern statistics and related tools provide additional avenues for robust, rank-based characterization that can mitigate sensitivity to calibration and mild nonstationarity \citep{Graffetal2013, ribeiro2017characterizing, unakafova2013efficiently}. Together, these developments enable a principled bridge between phenomenology in cosmic-ray series and generic indicators of dynamical change.

Against this backdrop, our contribution is to develop and evaluate a systematic pipeline that anticipates FDs using temporal markers derived from topological--geometric invariants computed on GCR time series across multiple stations and two FD episodes. We compute sliding-window estimates of entropy-, fractal-, and complexity-based measures, form independent marker series, and quantify station-wise temporal leads in minutes by aligning marker derivatives to a derivative-defined reference time of the original counts. We further report the fraction of stations that exhibit sustained pre-onset excursions under a fixed robust threshold-and-persistence rule, enabling operational comparison and selection of a compact marker panel within a unified feature framework \citep{fulcher2013highly, fulcher2014highly, lubba2019catch22, raubitzek2021combining, amigo2022information, Balasisetal2023, Gil2021EntropyKatz}.

Finally, because FD onsets reflect changes in transport conditions and effective diffusion barriers, they are natural testbeds for precursor discovery. The working hypothesis is that invariant-based markers—capturing changes in complexity, roughness, memory, or dimensionality—will exhibit detectable excursions ahead of amplitude drops when tracked with windows commensurate with interplanetary-disturbance scales. We therefore translate descriptive invariants into \emph{operational, station-level predictors} of FD onsets via a robust, threshold-and-persistence rule on pre-onset windows, integrating insights from nonlinear time-series analysis, comparative feature learning, and recent space-physics reviews \citep{Balasisetal2023, sierra2022fractal, fulcher2013highly, fulcher2014highly, lubba2019catch22, unakafova2013efficiently, amigo2022information, Raubitzeketal2021, zhou2014multifault, Stolzetal2017, ribeiro2017characterizing, Manisetal2021, strydom2021svd, Pandaetal2023, tao2018permutation, 2016arXiv161106190K, 2013arXiv1310.3564H, Graffetal2013}.

\section{Data}
\label{sec:data}

We use 1-minute galactic cosmic ray (GCR) count rates from the Neutron Monitor Database (NMDB) for two Forbush decrease (FD) episodes: 23 April 2023 (2023-04-23) and 10 May 2024 (2024-05-10). These two cases were selected to represent distinct FD morphologies while retaining broad multi-station coverage at the native 1-minute cadence.

For 2023-04-23, a coronal mass ejection (CME) impacted Earth and produced a clear global GCR depression followed by a gradual, multi–day recovery. The 2024-05-10 event was triggered by the shock of an interplanetary CME (ICME) and ranks among the largest recent FDs (up to about 15.7\% at $\sim$10~GV rigidity), coinciding with a severe geomagnetic storm and followed by a ground–level enhancement (GLE~74) on 11 May 2024 during the recovery phase \citep{Abunina2025arXiv,Papaioannou2025GLE74}.

For each event we analyze a curated subset of $N_s=28$ stations with continuous 1-minute coverage across the analysis window (see Methods for the continuity criteria). Minute-resolution counts are used in their station-specific native units as provided by NMDB. Downstream analyses operate on sliding-window invariant series computed from these counts; each invariant series is then standardized within station prior to derivative-based lead extraction, so that timing is assessed relative to each station's own baseline. For each event, the complete set of stations we used for the analysis is: 20230423: [ATHN, DOMB, SOPO, SOPB, THUL, NAIN, INVK, FSMT, TXBY, NRLK, APTY, OULU, CALG, DOMC, YKTK, KERG, NEWK, MXCO, NANM, KIEL2, BKSN, JUNG, TERA, LMKS, IRK2, IRKT, JUNG1, PWNK]; 20240510: [MXCO, DOMB, SOPO, SOPB, MWSN, MWSB, THUL, PWNK, NAIN, TXBY, APTY, OULU, DOMC, KERG, CALG, KIEL2, NANM, CALM, BKSN, JUNG, JUNG1, YKTK, LMKS, NEWK, TERA, IRK3, DRBS, INVK]. The short names of the stations can be found in \url{https://www.nmdb.eu/nest/} y \url{https://www.nmdb.eu/nest/help.php#helpstations}.

\begin{figure*}[htb]
\centering
\includegraphics[width=0.98\linewidth]{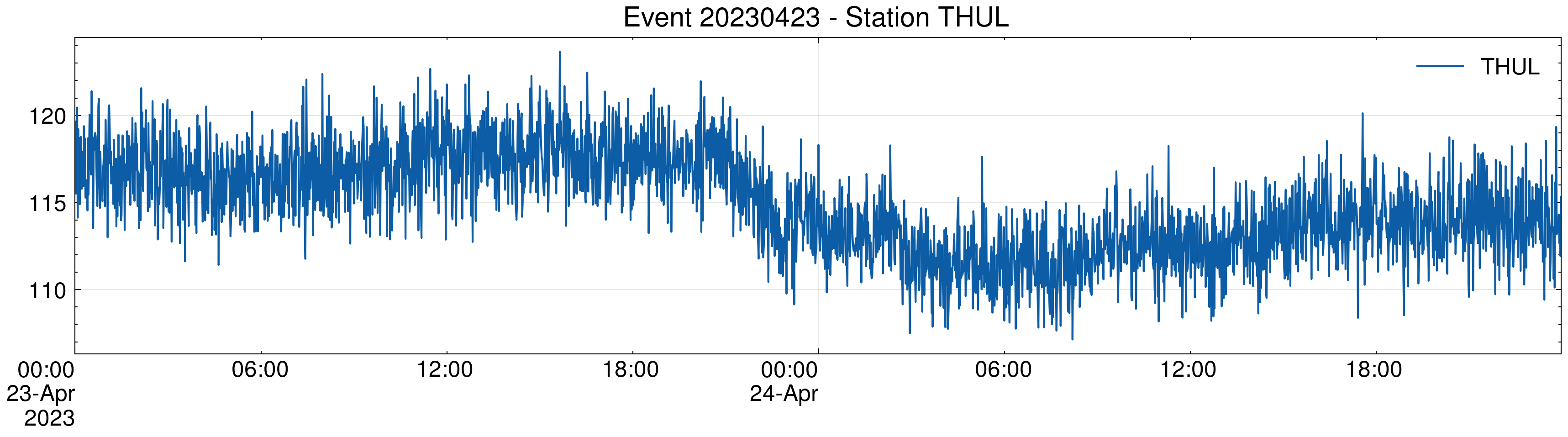}\\
\includegraphics[width=0.98\linewidth]{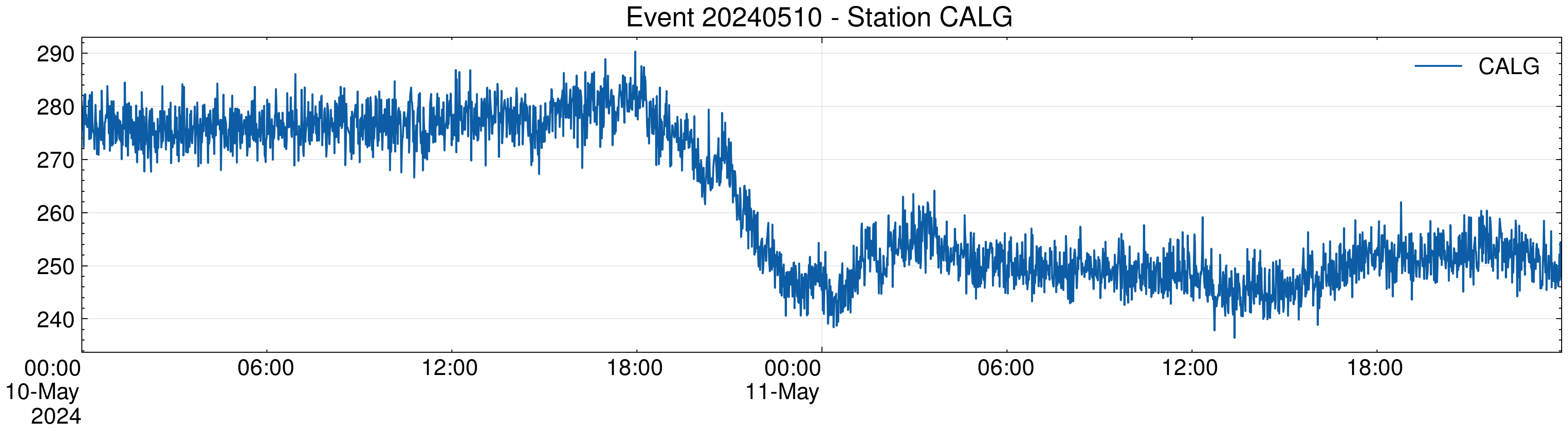}
\caption{Representative 1-minute NMDB count-rate series for the two analyzed FD events: THUL (2023-04-23) and CALG (2024-05-10). The panels illustrate distinct FD morphologies (a moderate event with gradual recovery versus an extreme event with large amplitude and a sharper main decrease), motivating a station-wise, derivative-aligned definition of the reference time used throughout the analysis.}
\label{fig:fd_examples}
\end{figure*}

Figure~\ref{fig:fd_examples} shows representative 1-minute count-rate series for each event (THUL for 2023-04-23 and CALG for 2024-05-10), highlighting the contrasting morphologies that motivate a station-wise, derivative-aligned alignment time.

\section{Methods}
\label{sec:methods}

\subsection{Data, smoothing, and operational alignment}
Let $x_s(t)$ denote the one-minute neutron-monitor count-rate series at station $s$ over an event-specific analysis window covering pre-event baseline, the main decrease, and early recovery. We retain stations with sufficient coverage in the immediate pre-onset interval and treat missing values conservatively by requiring a minimum valid fraction per analysis window (see sensitivity ranges below).

To reduce high-frequency noise while preserving abrupt transitions, we apply an exponentially weighted moving average (EWM) with coefficient $\alpha$,
\begin{equation}
\hat{x}_s(t) \;=\; \alpha\, x_s(t) \;+\; (1-\alpha)\,\hat{x}_s(t-1),
\label{eq:ewm_counts}
\end{equation}
initialized at the first time point of the event window. All subsequent computations operate on $\hat{x}_s(t)$.

Rather than adopting a threshold-based physical FD onset definition, we define a \emph{station-wise operational alignment time} $t_0^{(s)}$ anchored in the count-rate dynamics. Specifically, we compute the EWM-smoothed first difference
\begin{equation}
\Delta \hat{x}_s(t) \;=\; \hat{x}_s(t)-\hat{x}_s(t-1),
\label{eq:firstdiff}
\end{equation}
and set
\begin{equation}
t_{0}^{(s)} \;:=\; \arg\min_{t \in \mathcal{B}} \Delta \hat{x}_s(t),
\label{eq:t0_def}
\end{equation}
where $\mathcal{B}$ is an event-specific search bracket chosen to include the main decrease phase. Throughout the manuscript, $t_0^{(s)}$ is used as an alignment reference to synchronize heterogeneous station profiles; it is not claimed to be a universal physical onset time.

\subsection{Sliding-window invariants (marker series)}
For each station $s$, we compute a set of sliding-window invariants on the smoothed counts $\hat{x}_s(t)$ using a \emph{causal} window of length $W$ minutes evaluated every $\Delta t$ minutes. For evaluation times $t_j$ on the grid $\mathcal{T}$, we define the causal window
\[
\mathcal{W}(t_j) \;=\; \{\hat{x}_s(u): u \in [t_j-W,\,t_j]\}.
\]
The marker (invariant) series for metric $k$ is then
\begin{equation}
M_k^{(s)}(t_j) \;=\; \mathcal{F}_k\!\big(\mathcal{W}(t_j)\big), \qquad t_j \in \mathcal{T},
\label{eq:marker_def}
\end{equation}
where $\mathcal{F}_k$ denotes the estimator for invariant $k$.

This causal construction avoids look-ahead and makes the marker time series compatible with online early-flagging. We initially compute a broader screening set spanning information-theoretic, geometric/fractal, and complexity/embedding families; the main results focus on a compact CORE panel selected for high detection coverage and stable lead distributions across events.

\subsection{Marker dynamics and lead definition}
To summarize dynamical changes in markers, we smooth each marker series with an EWM (same $\alpha$ unless noted) and compute its first difference on the evaluation grid,
\begin{equation}
\Delta M_k^{(s)}(t_j) \;=\; \tilde{M}_k^{(s)}(t_j) - \tilde{M}_k^{(s)}(t_{j-1}),
\label{eq:marker_diff}
\end{equation}
where $\tilde{M}_k^{(s)}$ is the EWM-smoothed marker series. We define the lead for a station--marker pair as
\begin{equation}
\ell_{k,s}^{*} \;=\; t_{k,s} - t_0^{(s)},
\label{eq:lead_def}
\end{equation}
where $t_{k,s}$ is the time of the most negative excursion of $\Delta M_k^{(s)}(t)$ within the pre-onset search window $[t_0^{(s)}-L,\;t_0^{(s)})$. Leads are reported in minutes; $\ell_{k,s}^{*}<0$ indicates anticipation (marker change precedes $t_0^{(s)}$).

\subsection{Operational detector and station-wise coverage}
In addition to lead timing, we quantify whether a marker exhibits a sustained pre-$t_0$ excursion via an operational detector on the pre-onset window $[t_0^{(s)}-L,\;t_0^{(s)})$. We compute a robust $z$-score of the marker difference series using the median and MAD on the same pre-onset interval,
\[
z_{k,s}(t) \;=\; \frac{\Delta M_k^{(s)}(t) - \mathrm{median}(\Delta M_k^{(s)})}{1.4826\,\mathrm{MAD}(\Delta M_k^{(s)})},
\]
and flag a station--marker pair if $|z_{k,s}(t)| \ge Z_0$ for at least $d$ consecutive evaluation steps (bilateral threshold with persistence). This yields a station-wise detection indicator and, for each invariant $k$, the detection coverage $\mathrm{Detect}_k[\%]$ across the station network.

\subsection{Invariant estimators and default parameters}
The CORE panel used in \S\ref{sec:results} includes: Shannon entropy, spectral entropy, sample entropy, approximate entropy, Lempel--Ziv complexity, Higuchi fractal dimension, Katz fractal dimension, and correlation dimension. We follow standard definitions and widely used implementations for entropy-type measures and related parameterizations \citep{unakafova2013efficiently,amigo2022information}, DFA/Hurst and scaling estimators where used in screening \citep{zhou2014multifault,Stolzetal2017}, and correlation dimension \citep{Farmer1982}. Unless otherwise noted: spectral entropy uses a Welch PSD with sampling frequency $sf=1$ (one sample per minute) and normalization; Higuchi uses $k_{\max}=10$; and Lempel--Ziv uses binary symbolization by median thresholding within each window.

\subsection{Sensitivity analysis}
We probe robustness to methodological choices by varying: window length $W \in \{90,120,180\}$ min; stride $\Delta t \in \{1,5\}$ min; EWM coefficient $\alpha \in \{0.10,0.15,0.25\}$; pre-onset window $L \in \{360,480\}$ min; robust threshold $Z_0 \in \{1.3,1.5\}$; and persistence $d \in \{3,5,10\}$ evaluation steps. All qualitative conclusions reported in \S\ref{sec:results} are stable across these perturbations. For clarity and reproducibility, Table~\ref{tab:defaults-core} summarizes the default pipeline parameters used to produce the results in \S\ref{sec:results}, together with the ranges explored in the sensitivity analysis.

\begin{table}[htb]
\centering
\small
\caption{Default pipeline parameters (with sensitivity ranges). The marker windows are \emph{causal} (no look-ahead) and evaluated every $\Delta t$ minutes.}
\label{tab:defaults-core}
\begin{tabular}{@{}l c c l@{}}
\toprule
\textbf{Parameter} & \textbf{Symbol} & \textbf{Default} & \textbf{Range in sensitivity} \\
\midrule
Window length (min) & $W$ & 120 & $\{90,\,120,\,180\}$ \\
Stride (min) & $\Delta t$ & 5 & $\{1,\,5\}$ \\
EWM smoothing & $\alpha$ & 0.15 & $\{0.10,\,0.15,\,0.25\}$ \\
Pre--$t_0$ window (min) & $L$ & 480 & $\{360,\,480\}$ \\
Threshold (robust $z$) & $Z_0$ & 1.3 & $\{1.3,\,1.5\}$ \\
Persistence (steps) & $d$ & 5 & $\{3,\,5,\,10\}$ \\
Min valid fraction per window & --- & 0.95 & $\{0.90,\,0.95\}$ \\
\bottomrule
\end{tabular}
\end{table}

%===================================
\section{Results and Evaluation}
\label{sec:results}

We apply the pipeline described in \S\ref{sec:methods} to two Forbush decrease (FD) episodes with uniform 1-minute cadence in the NMDB station network: 2023-04-23 and 2024-05-10. For each event we analyze a 28-station subset and compute sliding-window invariants using a causal window ($W=120$ min) evaluated every $\Delta t=5$ min, so that each marker value $M_k(t)$ depends only on measurements up to time $t$ (no look-ahead). Throughout this section, the alignment time $t_0$ is defined operationally as the timestamp where the exponentially weighted moving-average (EWM) first difference of the smoothed count rate reaches its minimum at each station. Marker ``leads'' are reported in minutes as $\ell^* = t_k - t_0$, where $t_k$ is the time of the most negative excursion of the marker derivative $\dot M_k(t)$ within the pre-onset search window; hence $\ell^*<0$ indicates that the marker change precedes $t_0$.

\begin{figure*}[htb]
\centering
\includegraphics[width=0.49\linewidth]{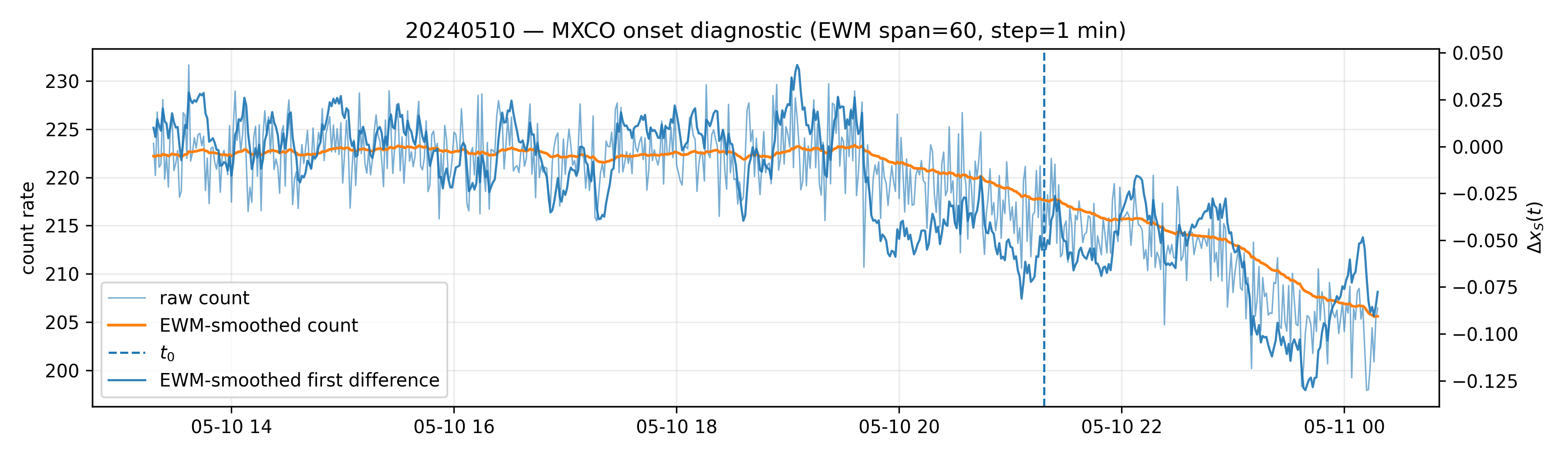}
\includegraphics[width=0.49\linewidth]{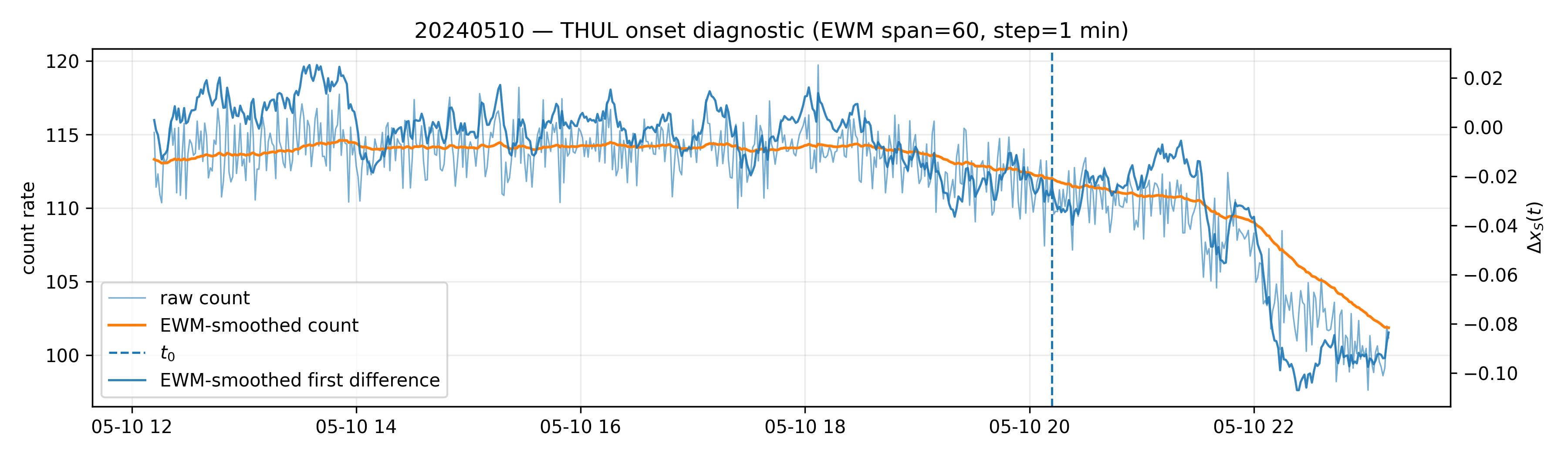}\\
\includegraphics[width=0.49\linewidth]{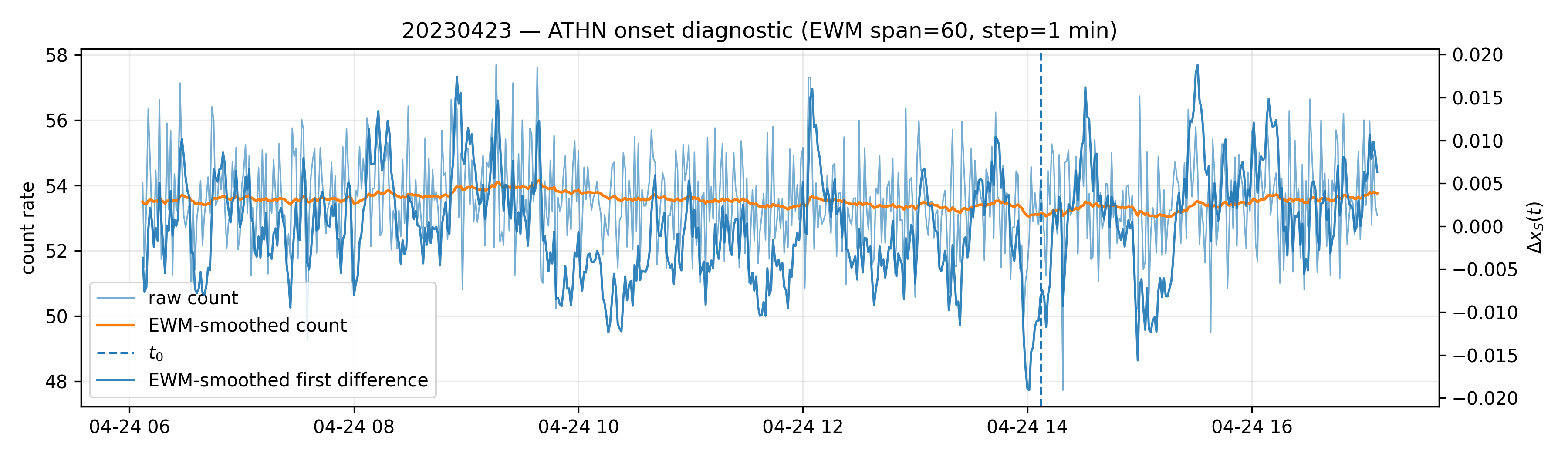}
\includegraphics[width=0.49\linewidth]{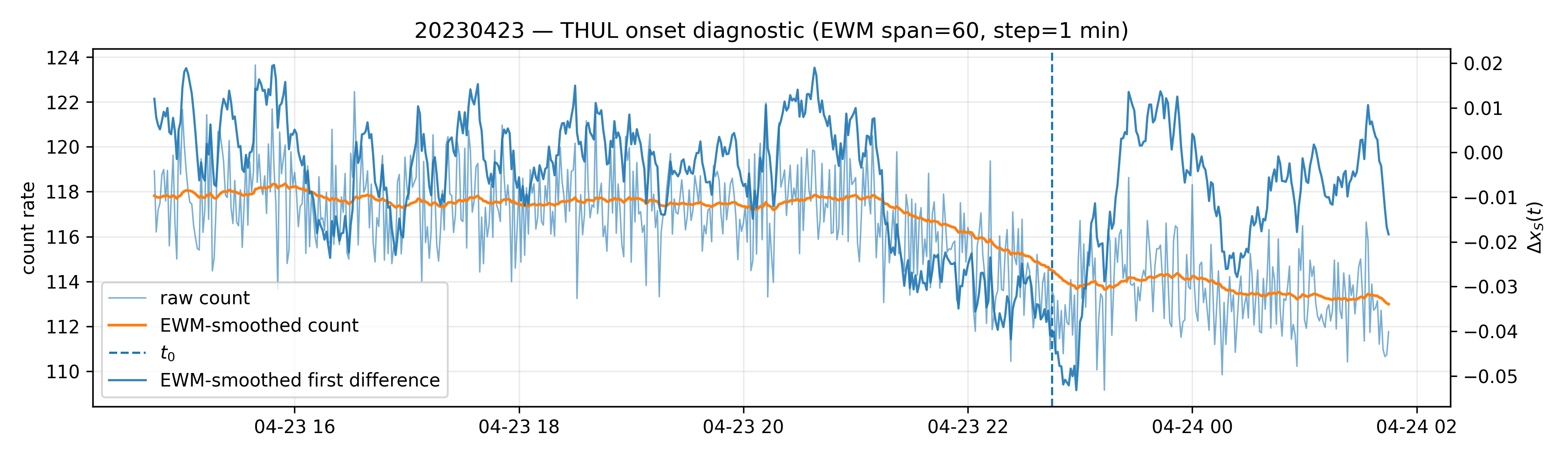}
\caption{Onset diagnostics for representative stations and events. For each panel, the raw NM count rate (light blue) is smoothed with an exponentially weighted moving average (EWM; orange), and the EWM-smoothed first difference $\Delta x_s(t)$ (dark blue, right axis) is used to define the reference onset time $t_0$ (vertical dashed line) as described in \S\ref{sec:methods}. This diagnostic illustrates the event morphology and the operational alignment used to measure marker leads/lags.}
\label{fig:onset_diag}
\end{figure*}

\paragraph{Onset alignment diagnostic.}
Figure~\ref{fig:onset_diag} illustrates the onset diagnostic on representative stations for each event, overlaying the raw count, the EWM-smoothed count, and the EWM-smoothed first difference. The selected $t_0$ consistently falls within the transition into the main decrease phase, at or near the strongest negative excursion of the smoothed first difference. We emphasize that $t_0$ is used here as an \emph{alignment reference} anchored in the count-rate dynamics, rather than as a threshold-based physical onset; this choice enables a station-by-station synchronization of heterogeneous profiles while preserving a fully automated definition that can be applied uniformly across stations and events.

\paragraph{A compact CORE panel.}
To reduce redundancy across closely related invariants, we focus the main evaluation on a compact CORE set spanning complementary families: information-theoretic measures (Shannon, approximate, sample, and spectral entropy), geometric/roughness descriptors (Higuchi and Katz fractal dimension), and complexity/embedding indicators (Lempel--Ziv and correlation dimension). Table~\ref{tab:core_ranking} reports pooled station-wise performance across both events, summarizing the fraction of station--event pairs flagged by the operational detector (Detect \%), and the distribution of station-wise leads. Across the two events, detection rates are consistently high (78--93\%), indicating that the CORE invariants frequently exhibit sustained pre-$t_0$ excursions under the robust threshold-and-persistence rule. Median leads are systematically negative and typically of order several hours: for example, spectral entropy shows a pooled median lead of $-365.5$ min (Detect 83.6\%), while Lempel--Ziv and Shannon entropy yield pooled medians of $-320$ min (Detect 81.8\%) and $-300$ min (Detect 92.7\%), respectively. These summaries motivate the CORE panel as a parsimonious operational set that retains the strongest and most repeatable anticipatory signatures while markedly reducing the methodological density of the full metric suite.

\begin{table}[htb]
\centering
\caption{CORE-panel station-wise lead statistics pooled across the two events (negative lead = precedes onset). 
Detect (\%) is the fraction of station--event pairs flagged by the operational detector.  $N$ is the number of station--event pairs with a valid lead estimate for that invariant (out of 56 total pairs; 28 stations $\times$ 2 events).}
\label{tab:core_ranking}
\begin{tabular}{lrrrrrr}
\toprule
Invariant & Detect (\%) & N & Median lead (min) & IQR (min) & Q25 & Q75 \\
\midrule
shannon\_entropy & 92.7 & 51 & -300.0 & 204.5 & -401.0 & -196.5 \\
app\_entropy & 90.9 & 50 & -203.0 & 197.0 & -308.5 & -111.5 \\
higuchi\_fd & 87.3 & 48 & -296.0 & 175.5 & -367.8 & -192.2 \\
sampen & 85.5 & 47 & -273.0 & 218.5 & -388.5 & -170.0 \\
corr\_dim & 85.5 & 47 & -227.0 & 221.0 & -340.0 & -119.0 \\
spectral\_entropy & 83.6 & 46 & -365.5 & 153.2 & -424.8 & -271.5 \\
lepel\_ziv & 81.8 & 45 & -320.0 & 187.0 & -432.0 & -245.0 \\
katz\_fd & 78.2 & 43 & -305.0 & 282.0 & -407.5 & -125.5 \\
\bottomrule
\end{tabular}
\end{table}

\begin{figure}[htb] 
\centering \includegraphics[width=0.9\linewidth]{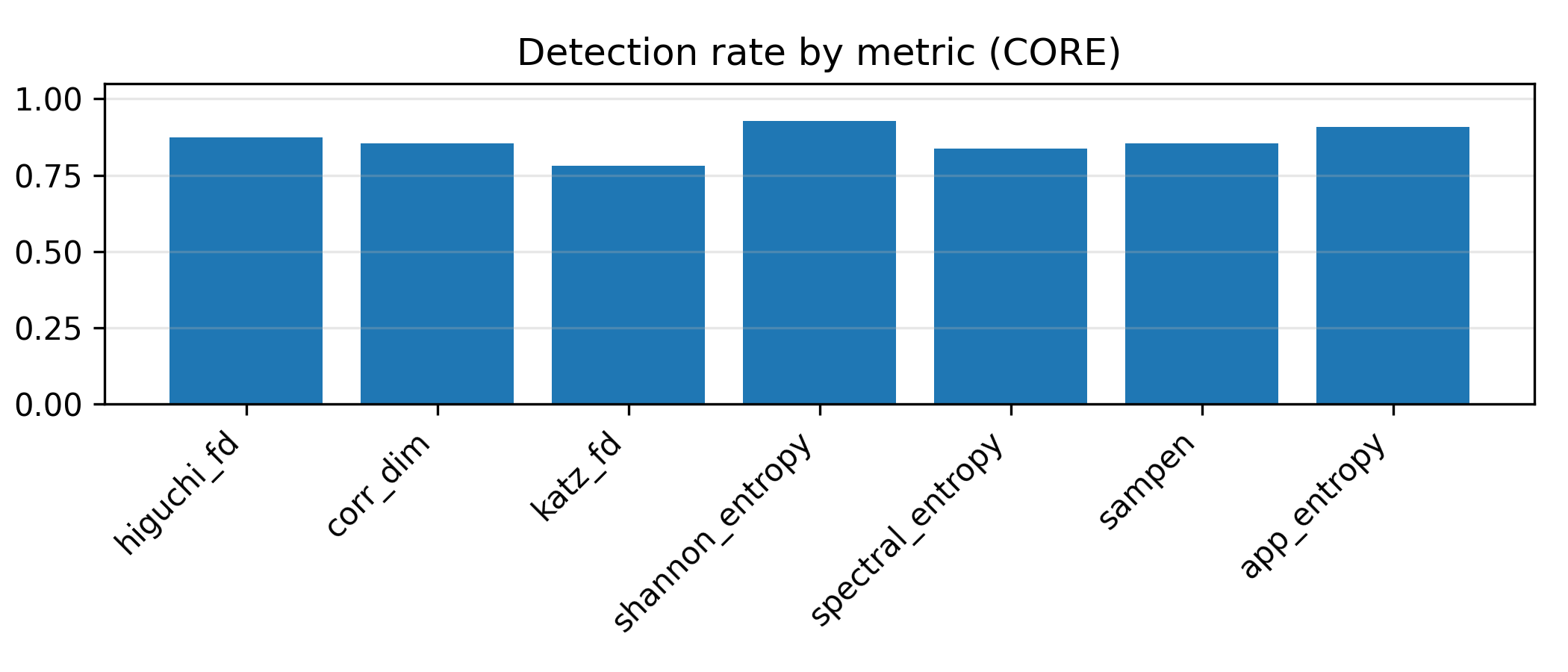} \caption{Detection rate by invariant for the CORE panel under the operational robust threshold-and-persistence detector (\S\ref{sec:methods}). Bars report the fraction of stations that exhibit sustained pre-onset excursions according to the predefined significance rule.} \label{fig:core_detection_rate} 
\end{figure}

\begin{figure*}[htb]
\centering
\includegraphics[width=\linewidth]{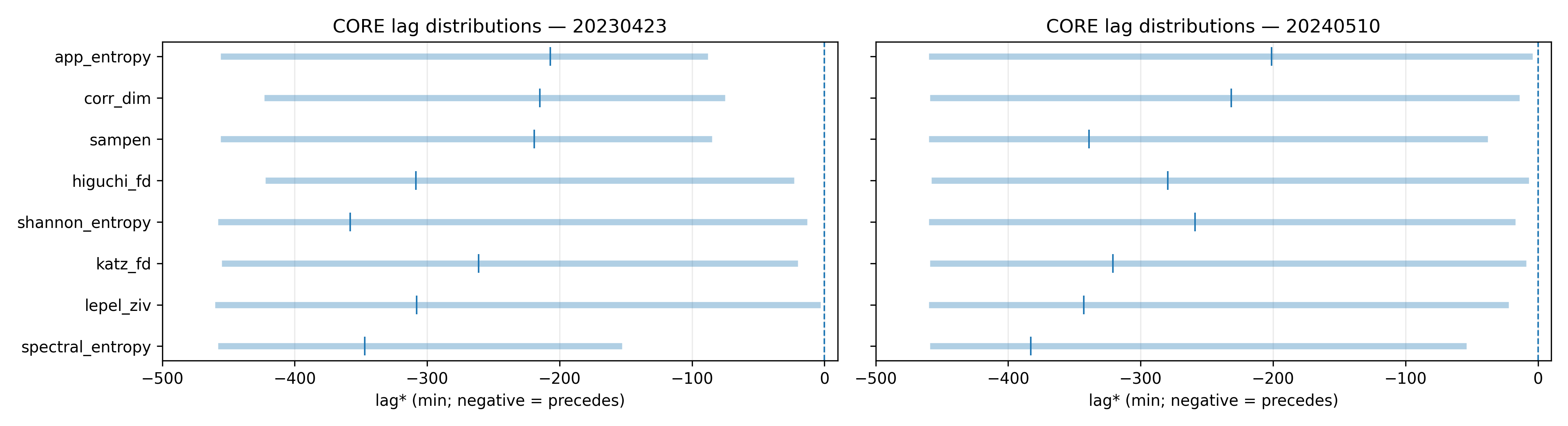}
\caption{Station-wise lag distributions for the CORE marker panel, shown separately for the two analyzed events. For each invariant, the horizontal segment summarizes the cross-station spread of the estimated lag $\ell^{\*}$ (in minutes) relative to the reference time $t_0$; negative values indicate anticipation (marker changes precede $t_0$). The vertical tick marks the station-level median lag for each invariant, and the dashed vertical line at $0$ marks contemporaneous response.}
\label{fig:core_lagbars}
\end{figure*}

\begin{figure*}[htb]
\centering
\includegraphics[width=\linewidth]{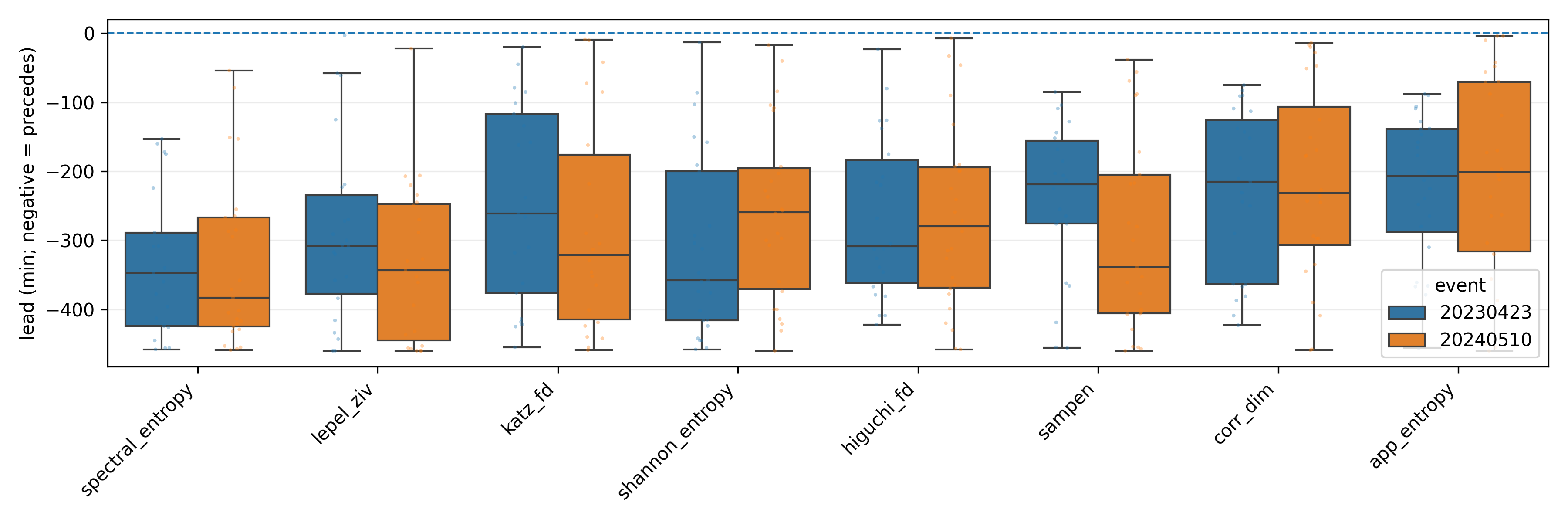}
\caption{Lead (lag) distributions by CORE invariant and event across the station network. Boxplots summarize the station-wise lags $\ell^{\*}$ (min) for each metric; dots show individual-station values. Negative values indicate anticipation (marker precedes $t_0$), and the horizontal dashed line at $0$ marks contemporaneous response. This view highlights both robustness (tight interquartile ranges) and heterogeneity (tails/outliers) across stations.}
\label{fig:core_boxpoints}
\end{figure*}

\paragraph{Detection coverage across invariants.}
Figure~\ref{fig:core_detection_rate} summarizes the detection coverage of each CORE invariant under the same robust threshold-and-persistence rule. The ordering is consistent with Table~\ref{tab:core_ranking}: most invariants trigger in a large majority of station--event pairs, indicating that the anticipatory signal is not confined to a small subset of stations. Importantly, this figure isolates \emph{coverage} from \emph{timing}: a metric may be highly detectable yet exhibit a shorter median lead, so Detect (\%) should be interpreted jointly with the lead distributions (Figs.~\ref{fig:core_lagbars}--\ref{fig:core_boxpoints}).

\paragraph{Station-wise lead dispersion and event dependence.}
Figure~\ref{fig:core_lagbars} shows the station-wise lead distributions for the CORE panel separately for each event using median ticks and interquartile ranges (IQR). The distributions reveal substantial station-to-station variability, with IQRs commonly spanning $\sim$2--4 hours depending on the invariant and event. This dispersion is expected given differences in geomagnetic cutoff rigidity, asymptotic directions, and local count-rate noise characteristics, and it reinforces the need for robust, station-level criteria rather than single-station inference. 

Complementarily, Figure~\ref{fig:core_boxpoints} overlays boxplots and individual station points for both events, making explicit both the central tendency and the prevalence of outliers (e.g., stations with near-zero or weakly negative leads) that contribute to the observed IQR.

\paragraph{Event-wise summaries.}
Table~\ref{tab:core_by_event} reports per-event CORE summaries (Detect \%, median lead, and IQR). For 2023-04-23, Shannon entropy exhibits a strong median lead of $-358$ min with 92.6\% detection, accompanied by spectral entropy at $-347$ min (77.8\%) and both Higuchi fractal dimension and Lempel--Ziv near $-308$ min (81.5\%). The 2024-05-10 episode displays a similar anticipatory pattern but with shifts in which invariants lead most strongly: spectral entropy reaches a median of $-383$ min (89.3\%), Lempel--Ziv $-343$ min (82.1\%), and sample entropy $-339$ min (89.3\%), while Shannon entropy remains highly detectable (92.9\%) with a more moderate median lead ($-259$ min). These differences are consistent with the fact that the events exhibit distinct decrease morphologies and station responses, yet the CORE panel maintains broad pre-$t_0$ detectability and predominantly negative lead distributions in both cases.

\begin{table}[htb]
\centering
\small
\caption{CORE-panel performance by event (station-wise). Detect (\%) is the fraction of stations flagged by the operational detector; leads are in minutes (negative = precedes $t_0$).}
\label{tab:core_by_event}
\begin{tabular}{l r r r r r r}
\toprule
& \multicolumn{3}{c}{\textbf{2023-04-23}} & \multicolumn{3}{c}{\textbf{2024-05-10}} \\
\cmidrule(lr){2-4}\cmidrule(lr){5-7}
\textbf{Invariant} & Detect (\%) & Median & IQR & Detect (\%) & Median & IQR \\
\midrule
shannon\_entropy   & 92.6 & -358.0 & 216.0 & 92.9 & -259.0 & 175.0 \\
app\_entropy       & 88.9 & -207.0 & 148.8 & 92.9 & -201.0 & 245.8 \\
higuchi\_fd        & 81.5 & -308.5 & 178.5 & 92.9 & -279.5 & 174.2 \\
sampen             & 81.5 & -219.0 & 120.0 & 89.3 & -339.0 & 201.0 \\
corr\_dim          & 85.2 & -215.0 & 238.0 & 85.7 & -231.5 & 200.0 \\
spectral\_entropy  & 77.8 & -347.0 & 135.0 & 89.3 & -383.0 & 158.0 \\
lepel\_ziv         & 81.5 & -308.0 & 142.8 & 82.1 & -343.0 & 198.0 \\
katz\_fd           & 77.8 & -261.0 & 259.0 & 78.6 & -321.0 & 238.5 \\
\bottomrule
\end{tabular}
\end{table}

\begin{figure*}[htb]
\centering
\includegraphics[width=0.49\linewidth]{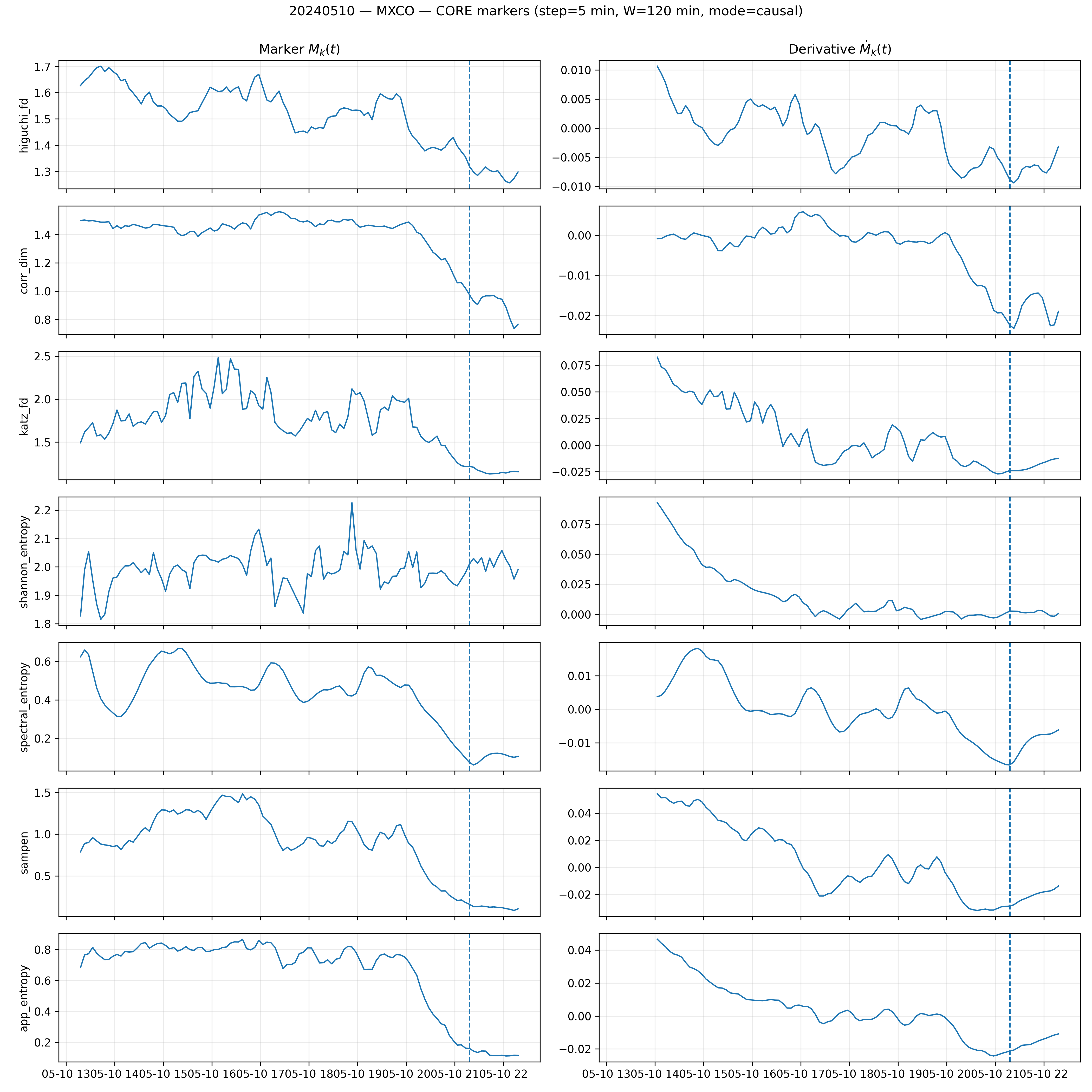}
\includegraphics[width=0.49\linewidth]{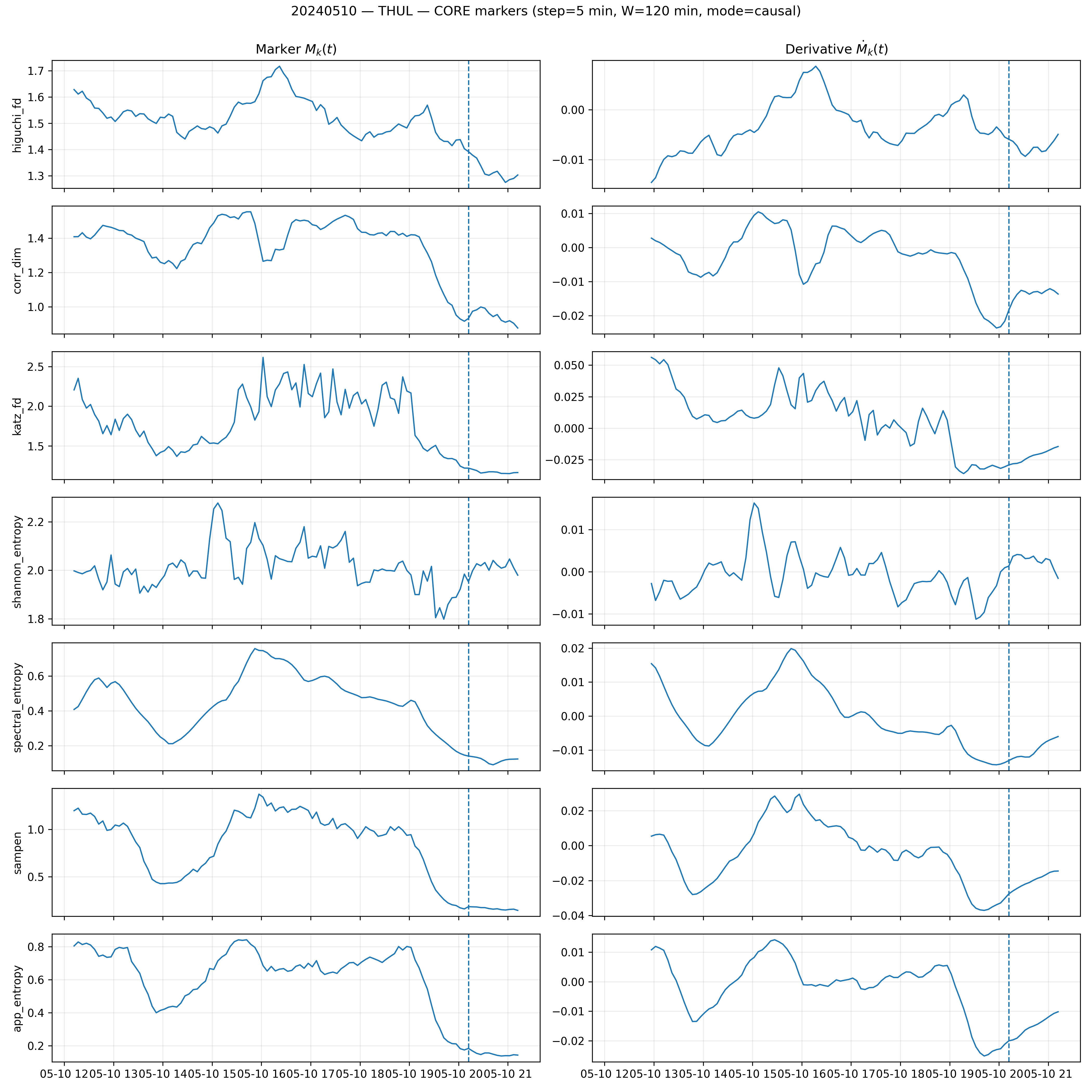}
\caption{CORE marker time series $M_k(t)$ (left column) and their EWM-smoothed derivatives $\dot{M}_k(t)$ (right column) for two representative stations during the 2024-05-10 event. The vertical dashed line marks $t_0$. The computation uses a causal sliding window (step and window length as annotated in the panels).}
\label{fig:markers_core_20240510}
\end{figure*}

\begin{figure*}[htb]
\centering
\includegraphics[width=0.49\linewidth]{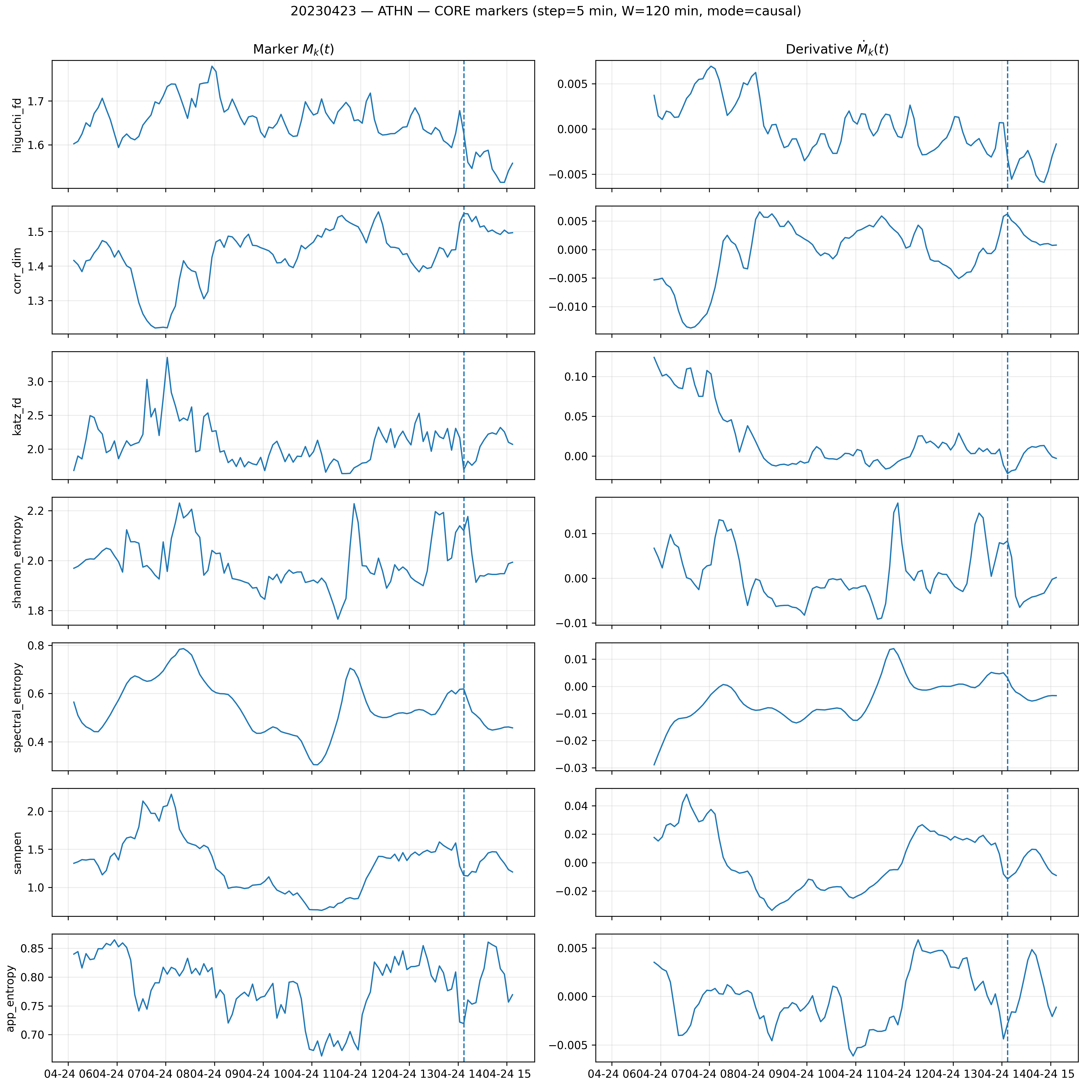}
\includegraphics[width=0.49\linewidth]{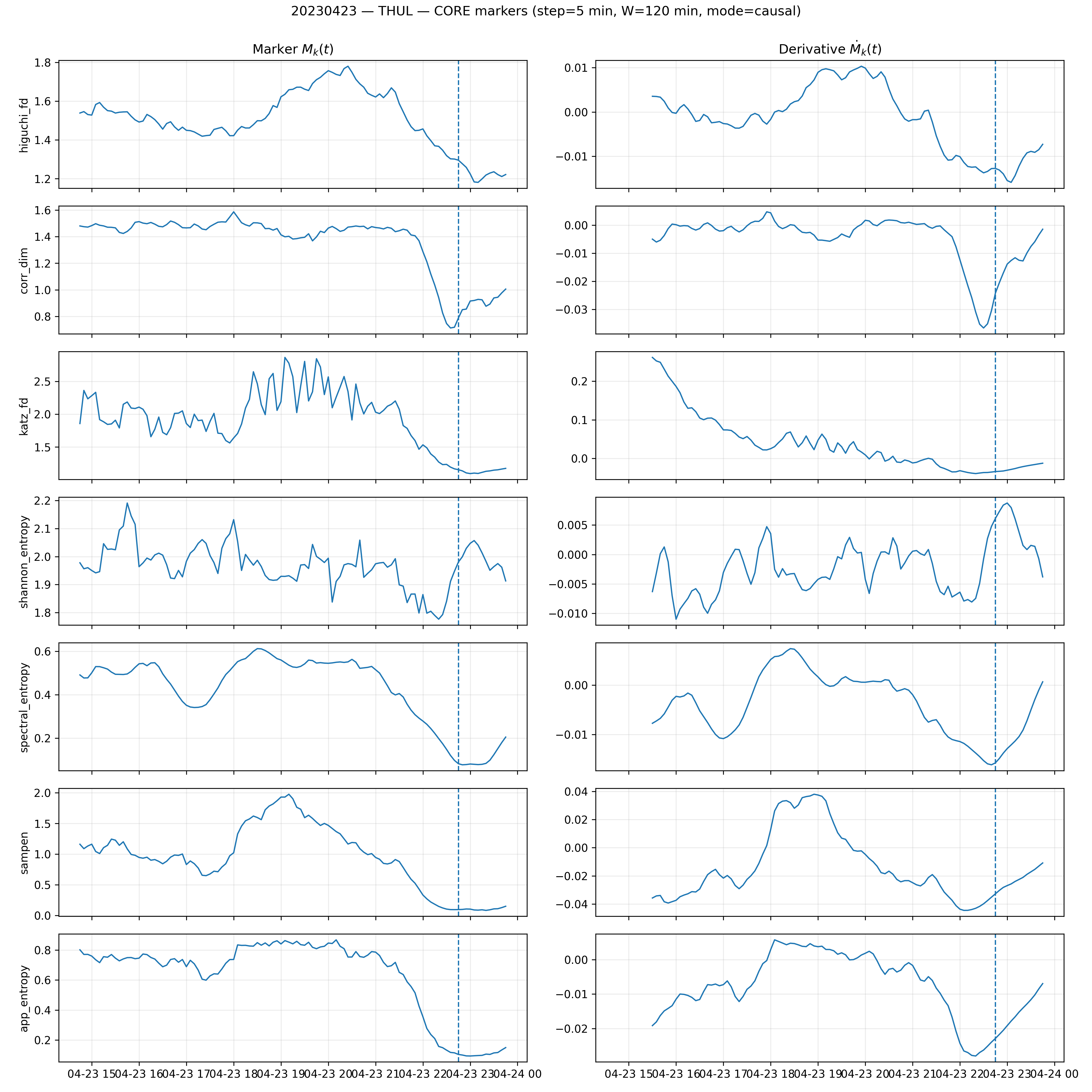}
\caption{Same as Fig.~\ref{fig:markers_core_20240510}, but for the 2023-04-23 event. These examples illustrate how anticipatory excursions appear as sustained departures in $\dot{M}_k(t)$ prior to $t_0$, while also showing station-dependent variability.}
\label{fig:markers_core_20230423}
\end{figure*}

\paragraph{Qualitative behavior of markers and their derivatives.}
To connect the summary statistics with the underlying time-domain behavior, Figure~\ref{fig:markers_core_20240510} and~\ref{fig:markers_core_20230423} present representative time series of $M_k(t)$ and $\dot M_k(t)$ for selected stations in each event. In the 2024-05-10 case, multiple CORE markers display coherent pre-$t_0$ structure: entropy-type measures and complexity/embedding indicators tend to decline into the decrease phase, while geometric descriptors show concomitant changes consistent with a temporary reduction in irregularity. The marker derivatives $\dot M_k(t)$ exhibit sustained negative excursions in the pre-$t_0$ window, providing the operational basis for early flagging. For 2023-04-23 the same qualitative organization is present but is more heterogeneous across stations, which is reflected in broader IQRs for several invariants in Table~\ref{tab:core_by_event} and Figure~\ref{fig:core_boxpoints}. Importantly, these examples supply the visual grounding requested by the referee: they demonstrate that the proposed lead extraction is tied to identifiable features in the marker trajectories rather than being an artifact of tabulated summaries.

\paragraph{Summary.}
Overall, the evaluation indicates that a compact CORE panel can provide reproducible, station-wise anticipatory signatures relative to the operational alignment time $t_0$, with median leads typically ranging from $\sim$3 to 6 hours and detection rates above $\sim$80\% for most CORE invariants. The combination of onset diagnostics (Figure~\ref{fig:onset_diag}), distributional summaries (Table~\ref{tab:core_ranking}, Figure~\ref{fig:core_lagbars}, Figure~\ref{fig:core_boxpoints}), and representative marker trajectories (Figure~\ref{fig:markers_core_20240510} and~\ref{fig:markers_core_20230423}) yields a complete and interpretable account of both the quantitative performance and the qualitative mechanisms underlying the anticipatory behavior.

\section{Discussion}
\label{sec:discus}

Across two FD episodes and a 28-station network subset, the results support a consistent operational picture: a compact set of information-theoretic, geometric, and complexity/embedding invariants exhibits sustained pre-$t_0$ structure with high station-wise detection coverage and predominantly negative lags. In particular, entropy-type measures (Shannon, spectral, sample, approximate), geometric descriptors (Katz and Higuchi fractal dimensions), and complexity/embedding indicators (Lempel--Ziv and correlation dimension) form a coherent panel that repeatedly departs from pre-onset baselines before the steepest decline in the smoothed count-rate dynamics. The fact that multiple families respond in a temporally ordered way, rather than a single metric dominating, mirrors broader findings that entropy/complexity and geometric signatures can capture regime shifts in space-physics time series and related geospace observables \citep{Koikkalainen2025NPG,Raath2022JGR,Gil2021EntropyKatz}.

A key implication is that the anticipatory behavior appears as a \emph{network} signature rather than a single-station artifact: detection fractions remain high across stations, while lead times exhibit substantial dispersion. Such heterogeneity is expected given differences in cutoff rigidity, asymptotic directions, and local noise characteristics, and is consistent with prior evidence that FD development and anisotropy can vary substantially across latitude bands and detector types during severe disturbances \citep{Riggi2025HLatFD}. Accordingly, the most robust operational use case is station-agnostic deployment with ensemble aggregation (e.g., requiring agreement across multiple stations and/or multiple markers) rather than relying on any individual station as definitive.

Event dependence is also informative. The 2024--05--10 episode shows particularly coherent anticipation across the CORE panel, consistent with its classification as an extreme heliospheric disturbance. Reports on the May 2024 storm period emphasize unusually strong FD behavior and associated high-energy particle activity (including the GLE~74 episode), providing context for why multiple statistical families might exhibit sustained departures over extended pre-$t_0$ intervals \citep{Abunina2025arXiv,Papaioannou2025GLE74}. In contrast, the 2023--04--23 event displays the same qualitative organization but with broader station-to-station spreads, reinforcing that FD morphology and observational geometry modulate both the timing and the clarity of precursory signatures \citep{Riggi2025HLatFD}. Importantly, our interpretation remains operational: we do not claim that any particular invariant uniquely identifies a physical microprocess; rather, we show that several complementary invariants reliably change \emph{before} the steepest count-rate decline, which is the key requirement for an early-flagging scheme.

Methodologically, two design choices matter for how these results should be read. First, $t_0$ is used as an \emph{alignment reference} derived from the count-rate dynamics (minimum of the EWM-smoothed first difference), not as a universal, threshold-based physical onset. This enables station-by-station synchronization of heterogeneous profiles without event-specific tuning. Second, the marker computation uses a causal window, which avoids look-ahead and makes the extracted leads compatible with online deployment. In this setting, the robust threshold-and-persistence detector provides a practical way to translate marker derivatives into reproducible timing summaries (coverage, medians, IQRs) without cross-correlation or formal hypothesis testing, while preserving interpretability of the underlying marker trajectories \citep{Koikkalainen2025NPG,Raath2022JGR}.

Several limitations remain. We use NMDB one-minute counts in native station units and do not attempt rigidity homogenization or additional response corrections beyond the data provider. Window length, step, and smoothing parameters necessarily trade temporal precision for noise suppression, and truly optimal settings may vary across event classes and station noise regimes. Moreover, although geometric and entropy/complexity markers are computationally light, some estimators (especially those tied to memory/scaling in broader screenings) can be sensitive in finite samples; recent work suggests that Bayesian or likelihood-based approaches to long-memory characterization may improve robustness at additional computational cost \citep{Borin2024PRE,Mangalam2025Algorithms}. Finally, while our results are consistent across two contrasting events, generalization requires larger multi-event evaluation across solar-cycle phases and rigidity bands, particularly under strong anisotropy.

Natural extensions follow. Multi-scale or adaptive windows could better accommodate both gradual precursors and abrupt transitions, and directionality could be handled explicitly for invariants whose excursions are not consistently signed across all morphologies. Multiscale generalizations of geometric descriptors (e.g., Katz-type constructions) are a promising route to increase sensitivity without prohibitive cost \citep{Li2023MSKFD}. Beyond univariate NM counts, combining the invariant panel with upstream solar-wind/IMF context could provide multimodal confirmation and improve alert confidence; recent entropy/complexity analyses of solar-wind structures support the diagnostic value of information-theoretic features for regime discrimination \citep{Koikkalainen2025NPG}. These steps would move the present case-study evidence toward calibrated, operational early-flagging in a network setting.

\section{Conclusions}
\label{sec:concl}

We presented and evaluated an invariant-based, derivative-aligned framework to anticipate Forbush decreases (FDs) in neutron-monitor count-rate records using a fully automated, station-wise procedure. The method converts EWM-smoothed count series into sliding-window invariants and extracts event-relative timing by aligning each station to an operational reference time $t_0$, defined as the minimum of the EWM-smoothed first difference of the count rate. Marker leads are then quantified in minutes from sustained excursions in the derivatives of the invariant time series within a pre-onset search window, and summarized through detection coverage and robust distributional statistics (medians and IQRs) across the station network.

Applied to two one-minute-cadence FD episodes (2023-04-23 and 2024-05-10) across 28 stations per event, the results show that a compact CORE panel spanning information-theoretic measures (Shannon, spectral, sample, and approximate entropy), geometric descriptors (Katz and Higuchi fractal dimension), and complexity/embedding indicators (Lempel--Ziv and correlation dimension) provides reproducible anticipatory signatures relative to $t_0$. Across the network, these invariants achieve high detection coverage (typically $\sim$78--93\%) and predominantly negative station-wise lags, with typical median lead times on the order of several hours. While lead distributions remain heterogeneous across stations, the combination of high coverage and consistent negative medians supports the interpretation that anticipatory structure is a network-level signal rather than a single-station artifact.

Finally, the onset diagnostic and the marker time-series examples demonstrate that the reported leads correspond to identifiable, persistent features in the invariant trajectories rather than isolated extrema. Together, the onset diagnostics, pooled/event-wise rankings, and station-wise distributions support the CORE panel as a parsimonious basis for early flagging in a multi-station setting. Future work should extend the evaluation to larger event sets spanning solar-cycle phases and rigidity bands, and explore adaptive/multiscale windows and multimodal fusion with upstream solar-wind and IMF context to improve calibration and operational alerting.

\section*{Data and Code Availability}

Minute–resolution neutron–monitor data were obtained from the Neutron Monitor Database (NMDB, \url{https://www.nmdb.eu}). In accordance with NMDB terms of use, we do not redistribute raw counts; the same station series for the Two events (2023-04-23 and 2024-05-10) can be retrieved directly from NMDB. All analysis code to reproduce the figures, tables, and statistics is openly available at \url{https://github.com/sierraporta/FD-characterization-topological-approach/tree/main}. For exact reproducibility, we recommend citing a tagged release or commit hash of the repository and recording the NMDB query metadata used to export the minute–resolution series.

\section*{Acknowledgments}

We gratefully acknowledge the Dirección de Investigaciones at Universidad Tecnológica de Bolívar for their support and accompaniment throughout this research process. We acknowledge the NMDB database (\url{www.nmdb.eu}), founded under the European Union's FP7 programme (contract no.\ 213007), for providing data.

\bibliographystyle{apsrev4-1}
\bibliography{main}

\end{document}